# Micron-Size Two-Dimensional Methylammonium Lead Halide Perovskites

*Eugen Klein[1], Andres Black[1], Öznur Tokmak[2], Christian Strelow[1], Rostyslav Lesyuk[1,3], Christian Klinke[1,4,5]\**

[1] *Institute of Physical Chemistry, University of Hamburg, Martin-Luther-King-Platz 6, 20146 Hamburg, Germany*

[2] *Fraunhofer Center for Applied Nanotechnology (CAN), Grindelallee 117, 20146 Hamburg, Germany*

[3] *Pidstryhach Institute for applied problems of mechanics and mathematics of NAS of Ukraine, Naukowa str. 3b, 79060 Lviv & Department of Photonics, Lviv Polytechnic National University, Bandery str. 12, 79000 Lviv, Ukraine*

[4] *Department of Chemistry, Swansea University – Singleton Park, Swansea SA2 8PP, United Kingdom*

[5] *Institute of Physics, University of Rostock, Albert-Einstein-Straße 23, 18059 Rostock, Germany*

**ABSTRACT**

Hybrid lead halide perovskites with 2D stacking structures have recently emerged as promising materials for optoelectronic applications. We report a method for growing 2D nanosheets of hybrid lead halide perovskites (I, Br and Cl), with tunable lateral sizes ranging from 0.05 to 8 µm, and a structure consisting of $n$ stacked monolayers separated by long alkylamines, tunable from bulk down to $n=1$. The key to obtaining such a wide range of perovskite properties hinged on utilizing the respective lead halide nanosheets as precursors in a hot-injection synthesis that afforded careful control over all process parameters. The layered, quantum confined ($n\leq4$) nanosheets were comprised of major and minor fractions with differing $n$. Energy funneling from low to high $n$ (high to low energy) regions within a single sheet, mediated by the length of the ligands between stacks, produced photoluminescent quantum yields as high as 49%. These large, tunable 2D nanosheets could serve as convenient platforms for future high efficiency optoelectronic devices.

\* Corresponding author: christian.klinke@uni-rostock.de

The outstanding properties of perovskites, including low temperature processability, tunable band gap,[1] small exciton binding energy,[2] narrow absorption edges and emission spectra, and long charge carrier diffusion lengths,[3] have been exploited in recent years for photovoltaic[4] and optoelectronic applications.[5] Bulk perovskites with the formula $ABX_3$ have been especially successful, where A is an organic ammonium cation or $Cs^+$, B is $Pb^{2+}$ and X is a halide anion. In contrast, 2D layered perovskite nanosheets have a Ruddleson-Popper type formula $L_2A_{n-1}B_nX_{3n+1}$, where L is a long chain alkyl[6] and *n* is the number of neighboring BX monolayers between the organic spacers. This crystal structure makes 2D perovskite nanosheets labile,[7,8] and for *n*<4 they are in quantum confinement. Like quantum wells, the low dielectric screening and large exciton binding energy in these quantum confined perovskites enhances their radiative recombination properties compared to bulk ($n=\infty$), whereas the long alkyl ligands enhances their stability.[9] Compared to films composed of small nanoparticles, individual 2D nanosheets do not exhibit tunnel barriers or grain boundaries in the lateral dimensions, which makes them interesting for optoelectronic studies[10,11] and flexible electronic devices.[12]

Perovskites nanosheets can be prepared through various approaches. Micron-sized sheets were obtained *via* a chemical vapor deposition growth method.[13] Colloidally stable methylammonium lead bromide (MAPbBr) and MAPbI nanosheets, up to 500 nm in size, can be prepared *via* exfoliation of bulk crystals with long chained ligands[14] or *via* instantaneous crystallization of precursor salts in an antisolvent.[15] Despite the significant progress made in the past years, synthesis of micron-size, colloidal, quantum-confined perovskites is limited to CsPbBr.[16] Such MAPbX nanosheets, where X is Cl, Br or I, have remained elusive.

We present in this work a colloidal method for producing MAPbX nanosheets for all halides, with control of sheet thickness and structure down to *n*=1, as well as lateral dimensions from <1 μm up to 8 μm. This hot-injection, colloidal synthesis method is based on dissolving as-prepared PbX nanosheets[17] and using them as precursor for the synthesis of the respective MAPbX nanosheets. The hot-injection route allows tuning the lateral dimensions and thickness of the nanosheets by controlling all aspects of the reaction process, from the ligand and precursor concentration, to the temperature and reaction times. In contrast, colloidal anti-solvent and exfoliation methods offer a much more limited control of the process parameters. The perovskite structures were prepared either as single crystal 2D particles or as stacked sheets consisting of repeatable $MA_{n-1}Pb_nX_{3n+1}$ layers separated by long ligands, as confirmed by X-Ray Diffraction (XRD). The layered structures, which were more stable due to the dense ligand shell,[9] exhibited quantum confinement, as confirmed by UV-Vis absorption and photoluminescent (PL) spectroscopy. Regions of different thickness within a single sheet produced an energy/charge carriers funneling phenomenon,[18] particularly strong in high PL quantum yield (PLQY) MAPbBr, resulting in emission from the lowest energy (highest $n$) sheets in the stack. The bottom-up synthesis approach for producing large area MAPbX nanosheets presented herein provides opportunities for both fundamental and applied optoelectronic research.

## RESULTS AND DISCUSSION

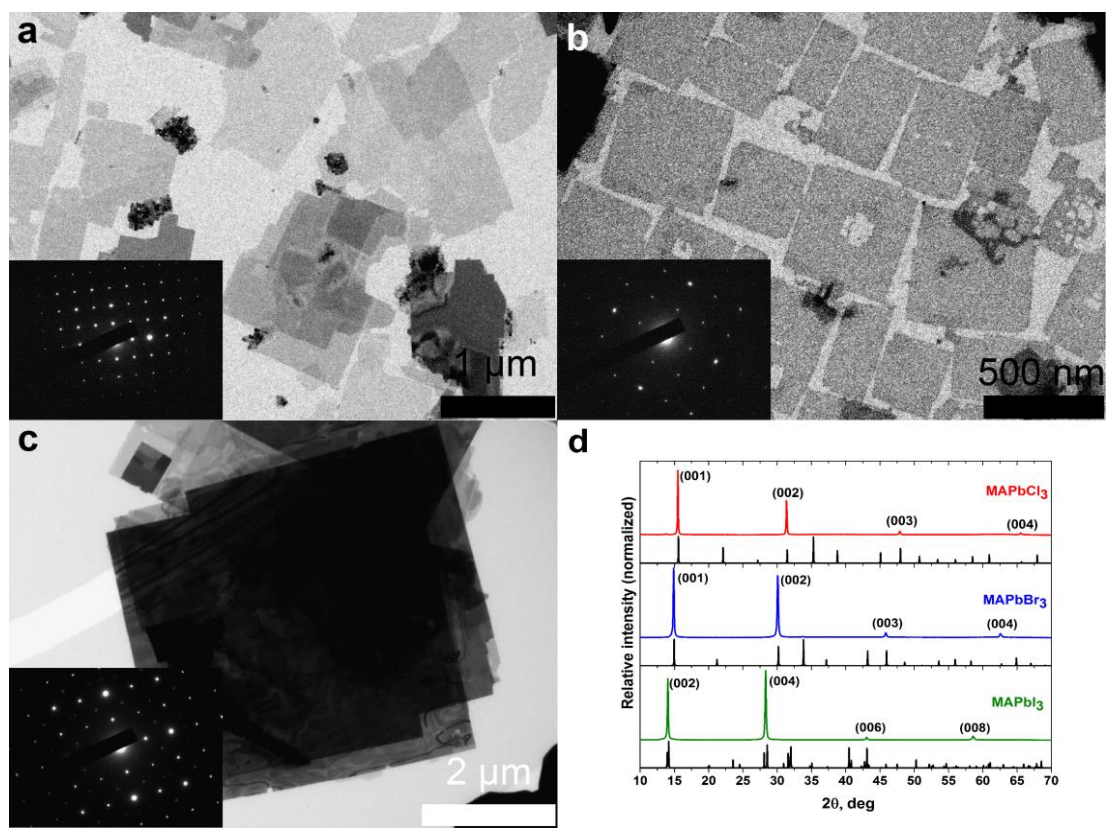

**Figure 1.** Shape and structural characterization of ($n>4$) MAPbX nanosheets. TEM images of (a) MAPbCl (b) MAPbBr and (c) MAPbI nanosheets. Insets show the SAEDs of the corresponding materials. (d), XRD patterns of the three structures, displayed for comparison with the reference diffraction patterns (black).[19–21]

Figures 1a, b and c present transmission electron microscope (TEM) images of bulk (*n*>4) nanosheets of MAPbCl, MAPbBr and MAPbI, respectively. All these structures were synthesized using the corresponding PbX nanosheets as precursors, shown in Figure S1. The MAPbCl nanosheets have straight edges, uniform thickness and lateral sizes from 500 nm to 1.5 μm. The MAPbBr nanosheets shown in Figure 1b are more uniform and square-like compared to the MAPbCl, with a size distribution between 400 and 600 nm. MAPbI nanosheets show the most uniform shape and size distribution, with the ones depicted in Figure 1c having lateral dimensions between 2 µm and 4 µm. Figures S2 to S4 show the wide range of tunable lateral sizes that can be prepared for the bulk nanosheets, and is summarized in Table 1. Whereas the structure of the MAPbCl changes from uniform squares to net-like with increasing size, the shape of the MAPbBr and MAPbI sheets is independent of their lateral size. Figure S5 depicts atomic force microscopy (AFM) images for all three materials. The thickness fits well with the data calculated from XRD patterns and can be tuned from 2.5 nm to 80 nm for MAPbCl, 2.6 nm to 50 nm for MAPbBr and 20 nm to 56 nm for MAPbI (Figure S6).

**Table 1**: Lateral size range of bulk and layered nanosheets.

| Material | Bulk (μm) | Layered (μm) |
|---|---|---|
| MAPbI | 0.05 – up to 8 | 0.05 - 6 |
| MAPbBr | 0.05 - 2 | 0.05 – up to 8 |
| MAPbCl | 0.3 – up to 8 | 1 - 15 |

The insets in Figure 1a-c presents selected area electron diffraction (SAED) patterns for all three materials, and confirm the single crystal appearance of the nanosheets. The lattice constants obtained from the SAED analysis correspond to a=b≈5.75 Å for MAPbCl NSs, a=b≈5.93 Å for MAPbBr NSs and a=b≈8.87 A for the MAPbI NSs. The appearance of only (*00l*) signals indicates that the lateral growth and the alignment of nanosheets (NSs) is orthogonal to the [001] direction. XRD patterns shown in Figure 1d reveal four signals for all three materials, which fit with (*00l*) reflections and confirm the orientation of the crystal lattice. The omission of the rest of the signals occurs due to the planar alignment of the sheets on the XRD wafer and the resulting texture effect. The lattice constant c =5.71 Å, 5.95 Å and 12.62 Å was determined from (*00l*) patterns for MAPbCl, MAPbBr and MAPbI NSs.

Capillary XRD (Figure S7) agrees with the reference spectrum, indicating that no other phases are present.

The bulk perovskite nanosheet structures with tunable lateral size and thickness were synthesized using the corresponding PbX nanosheets prepared in nonanoic or oleic acid (detailed protocols and instructions in the SI). The perovskite syntheses for all three materials employ diphenyl ether (DPE) as the solvent and trioctylphosphine (TOP) together with long chained amines as ligands. These chemicals were mixed, heated up to 80 °C and dried under vacuum for 1 h. The lead halide nanosheets in toluene were injected at temperatures between 80 °C and 220 °C and stirred until all material was dissolved. The syntheses were started by the addition of the methylammonium halides in dimethylformamide (DMF). The lateral size can be tuned for the MAPbCl by varying the concentration of the reactants, and for MAPbBr with the amount of ligands. For MAPbI, large sheets could be obtained by growing them slowly *via* a controlled temperature increase, from 35 to 110 °C, and by decreasing the amount of methylammonium iodide. In general, starting a reaction at high temperatures with a ratio of the reactants nearly at 1:1 leads to many nuclei that grow into small particles. In contrast to this, a slow increase in reaction temperature and a higher ratio between the two reactants of 5:1 leads to a reaction mixture consisting of few nuclei that grow into big structures.[22,23] The thickness of the bulk sheets can be increased by increasing the amount of methylammonium precursor for the MAPbCl and MAPbI nanosheets, and by increasing the concentration of the two reactants for MAPbBr. Aliquots taken during the reaction (Figure S8) show an agglomeration of small three dimensional particles arranged in the same size and shape as the end product. SAED image of these structures are comprised of dot pattern that indicates that the particles are all oriented in the same way and form quasi-crystal-units with crystal lattice similar to MAPbI nanosheets. These findings indicate that the sheets are formed either from small particles which agglomerate and merge, or that the precursors meet in some sort of micelle in the shape and size of the end product prior to nucleating, similar to PbS nanosheets[24] and in contrast to the continuous growth mechanism of different colloidal materials.[25,26]

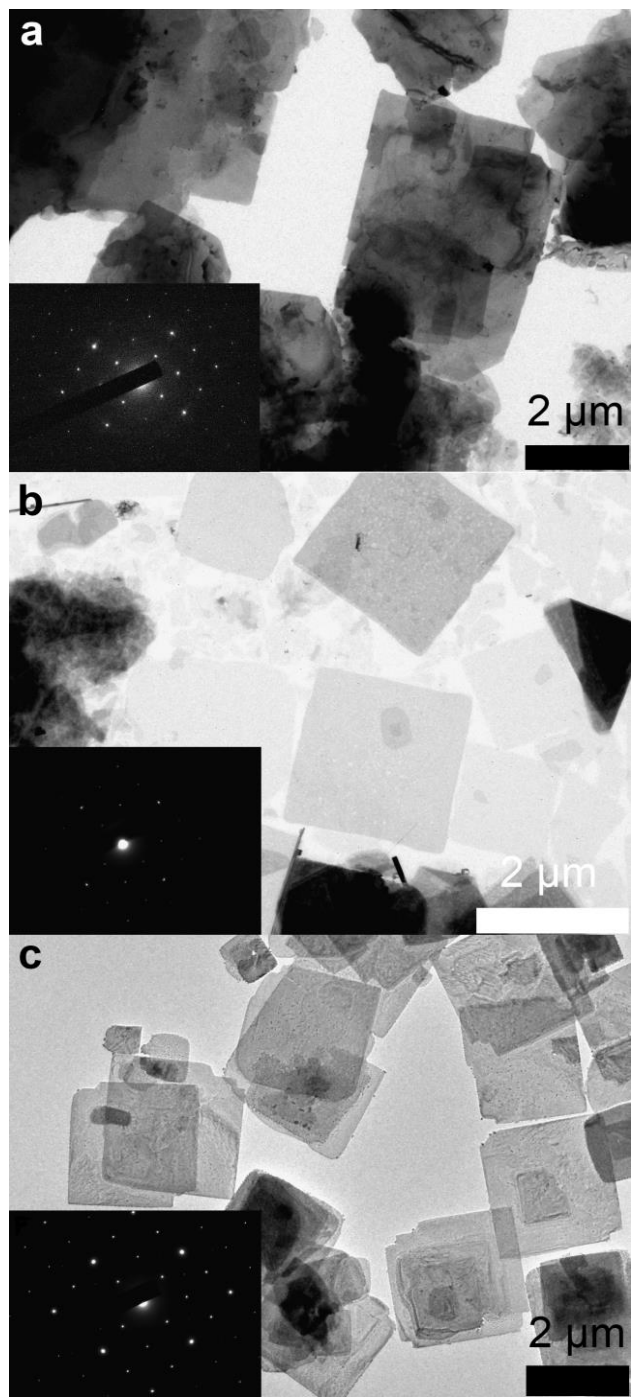

**Figure 2.** Shape characterization of layered ($n≤4$) nanosheets. TEM images of (a) MAPbCl, (b) MAPbBr and (c) MAPbI nanosheets. Insets show the SAEDs of the corresponding materials.

Figures 2a, b and c present TEM images of layered 2D nanosheets of MAPbCl, MAPbBr and MAPbI, respectively. The wide range of lateral tunability is presented in Figures S9 to S11, and summarized in Table 1. The MAPbCl sheets in Figure 2a show a large variation in shape

and lateral size, between 1 and 4 µm, whereas the MAPbBr sheets have a uniform square-like shape and a smaller size distribution, between 1 and 2 µm. Similar to their bulk counterparts, layered MAPbI nanosheets show the most uniform shape and size distribution, between 1.5 and 2.5 µm. Some MAPbBr and MAPbI layered nanosheets show pyramidal, shifted pyramidal or squared spiral stacking, as shown in Figure S12. All of the layered nanosheets produce a dot SAED pattern (Figure 2 insets), evidence of their monocrystallinity, corresponding to a cubic or tetragonal lattice of MAPbX viewed from [001].

The general synthesis of layered and bulk nanosheets is similar. The critical difference for obtaining layered sheets lies in utilizing a shorter alkyl chain ligand like hexadecylamine (HDA), which favors layer stacking, and lower reaction temperatures, ensuring the stability of the stacks. The lateral size of the MAPbCl sheets can be tuned by changing the amount of amine ligands, whereas the amount of both TOP and amine ligands determined the size of the MAPbI sheets. In contrast, careful control of the nucleation event is required to determine the size of the MAPbBr sheets. Injecting the methylammonium bromide precursor into the reaction mixture at 160 °C resulted in the immediate formation of 2 µm nanosheets. Maintaining the temperature for 5 minutes completely dissolved the structures. Finally, letting the solution cool slowly to 60 °C over 35 minutes leads to a nucleation around 80 °C and sheets of 4 to 8 µm in size. In general, the variation of lateral sizes of MAPbX nanosheets presented is constituted by several factors and is different for MAPbI, MAPbBr, MAPbCl, respectively. In some cases, two or three parameters are responsible for the lateral growth such as temperature and amount of ligand. Detailed protocols can be found in the SI. The variation in thickness can be controlled for the chloride sheets with the chain length of the amine, for bromide with the amount of the two reactants and for iodide with the temperature.

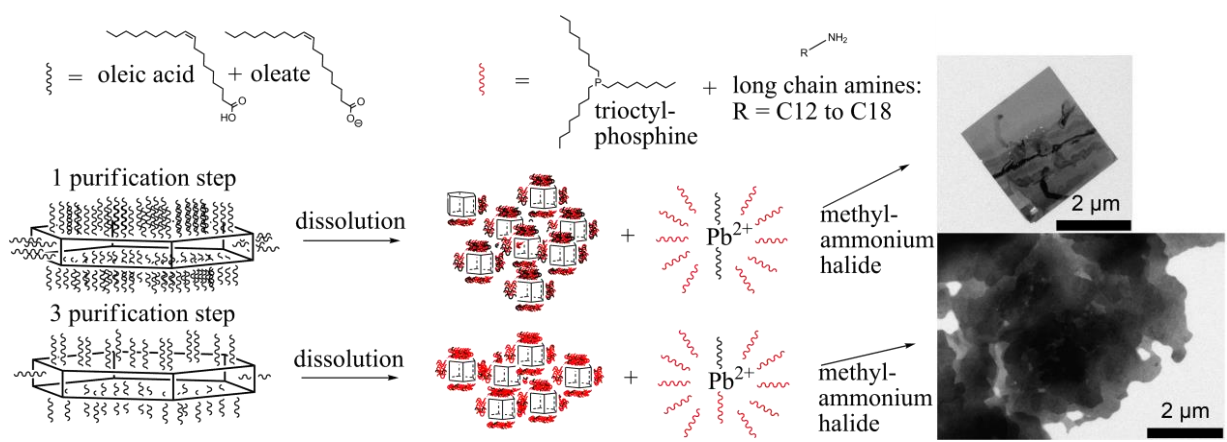

**Figure 3.** MAPbX synthesis. Schematic illustration depicting the importance of the precursor and its purification, using PbI nanosheets as an example.

Maintaining the surface ligands of the PbX precursor was critical for a successful synthesis. Therefore, as prepared PbX nanosheets were used, and were only centrifuged once, as shown in the center row of Figure 3, showing a schematic of the general synthesis process. In order to prepare MAPbX nanosheets, PbX nanosheets were dissolved in DPE along with TOP and long chained amine ligands. In the case of PbI, the mixture turns from turbid yellow to a slightly yellow pellucid solution. After dissolution, the resulting dissolved PbX nanoparticles and $Pb^{2+}$ ions were surrounded by a mixture of the original oleate and oleic acid ligands, and the newly added TOP and amines. UV/VIS and PL spectra, shown in Figure S13, reveal pronounced absorption and emission features in the range 330–400 nm which we attribute to tiny lead iodide nanoparticles with sizes below 1 nm. These eventually grew into the well-defined MAPbX perovskite nanosheets. The synthesis did not work if PbX powders were used, while excessive centrifugation (3 times) of the as prepared PbX nanosheets removed many of their ligands and produced amorphous structures with undefined shapes, as shown at the bottom right of Figure 3.

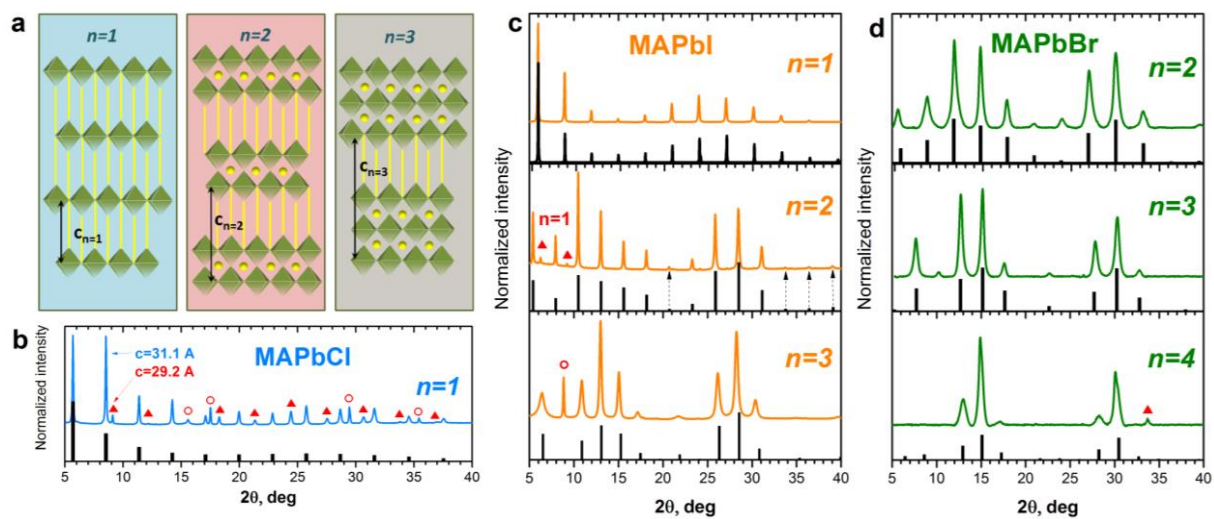

**Figure 4.** Structural characterization of layered MAPbX. (a) Schematic of layered MAPbX nanosheets for $n$=1,2,3. Organic part of the structure illustrated schematically in yellow. (b) Experimental (color) and simulated (black) XRD patterns for various $n$ for (b) MAPbCl, (c) MAPbI and (d) MAPbBr. Red triangles correspond to the presence of a minor fraction with different $n$, and red circles to a fractions coming from regions with different stacking or bulk domains. Enlarged patterns of Bragg reflections for corresponding structures can be found in Figure S14.

XRD, optical absorption and photoluminescent spectroscopy, shown in Figure 4 and 5, respectively, were used to study the crystallographic and structural properties of the MAPbX nanosheets, along with their resulting quantum confinement. Figure 4a depicts schematically the PbX nanosheets, consisting of $n$ perovskite monolayers between organic ligand spacers. Due to the strong texture effect, the powder XRD patterns of these samples (prepared by drop-casting diluted solutions, ensuring the lateral alignment of the sheets on the substrate) consists of repeatable equidistant (*00l*) Bragg reflections, whose relative intensities are determined by $n$. These spectra agree well for all materials and thicknesses with XRD patterns calculated from the respective inorganic PbX crystal structures for $n$=1, 2 and 3. Capillary XRD measurements, shown in Figure S15, eliminate the texture effects and confirm the perovskite structure of the layered sheets, and the half-unit cell shift between adjacent $n$-monolayer stacks within the same sheet, as reported previously for 2D Ruddleson-Popper perovskite crystals.[6] The XRD patterns reveal that the multi-layered nanosheets tend to form mixtures of different $n$ values, as indicated by the red triangles in the figures for the minor $n$ fraction. Relatively pure samples for different $n$ values were obtained for both the MAPbI and the MAPbBr based nanosheets, indicating the robustness and tunability of the synthesis process for these systems. The Bragg reflections positions were used to calculate the unit constant $c$, corresponding to the spacing between adjacent $n$-monolayer stacks, and subsequently the thickness of the organic layer spacer between the stacks. For MAPbI nanosheets, $c$=29.5 ($n$=1), 34.5 ($n$=2), and 40.6 Å ($n$=3), with an organic spacing between 2.1 and 2.3 nm, which agrees well with the 2.17 nm length of the HDA employed as ligand. Likewise, for MAPbBr, $c$ equals 29.6 ($n$=2), 35.4 ($n$=3), and 41.1 Å ($n$=4), and the organic layer spacing of 1.7 nm agrees well with the length of the dodecylamine (DDA) ligand. In both cases, the ligands between $n$-layer stacks appear to be interdigitated along the [001] axis, as illustrated in Figure 4a. Only $n$=1 stacking was obtained in relatively pure form for the MAPbCl nanosheets, with $c$=31.1 Å and an organic layer spacing of 2.5 nm, significantly longer than the expected 1.9 nm of the tetradecylamine (TDA) ligand. This indicates that the ligands between neighboring MAPbCl stacks are not fully interdigitated. The residual intensity of narrow non-Bragg reflexes for a given structure (for example in Figure 4b, c as indicated by the red circle) might originate from bulk regions within the sheets or regions with a more complex built-up.

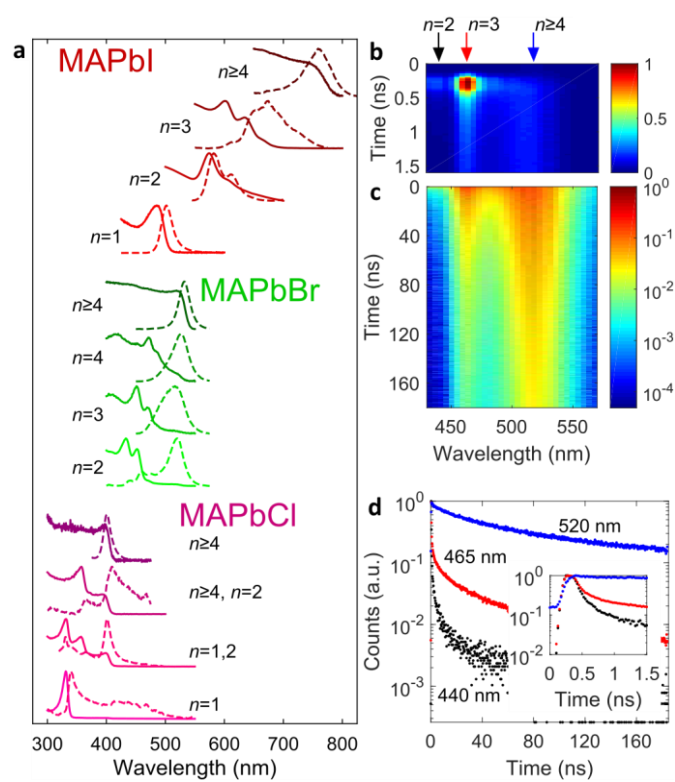

**Figure 5.** Optical properties of MAPbX nanosheets. (a) UV/VIS absorption and photoluminescence (PL) spectra (solid and dashed lines, respectively). Time-dependent PL decay of MAPbBr at various wavelengths, for (b) short (linear scale) and (c) long times (logarithmic scale). Arrows in (b) correspond to PL decay illustrated in (d) for specific wavelengths corresponding to different $n$.

The presence of regions of different thicknesses within the same nanosheet batch is supported by absorption and PL spectroscopy, from the near infrared for the MAPbI series to the near UV for the MAPbCl, as shown in Figure 5a. As the sheets transition from bulk to quantum confined, the absorption spectra reveal the emergence of sharp excitonic features.[14,27] The MAPbI $n=1$ sample shows excellent spectral purity, with a single absorption peak at 492 nm, and a slightly offset PL peak at 502 nm. Likewise, the bulk spectrum shows an absorption edge peaking at 730 nm, and PL at 759 nm. In contrast, the other two spectra in the MAPbI series show mixed $n$ fractions: the $n=2$ sample shows absorption at 574 nm and a shoulder at 600 nm, corresponding to the main $n=2$ and a minor $n=3$ fractions, with the respective PL peaks Stokes-shifted by about 10 nm. The main fraction $n=3$ spectrum also shows minor absorption features at longer wavelengths, with almost all of the PL coming from the thickest minor fraction at 674 nm.

In the MAPbBr series, the main PL peak for all of the samples is centered between 515 and 525 nm, at significantly lower energy than the exitonic absorption peaks arising from quantum confinement of the nanosheets. This PL emission at lower energy is likely due to an energy cascade mediated by excitonic charge transfer from high energy, quantum confined regions to lower energy bulk regions within the same nanosheets.[28] This is also known as energy funneling.[29] In this process, more confined, low $n$ regions with larger bandgap transfer their energy *via* resonant energy or charge transfer processes to less confined, higher $n$ regions with lower bandgap, where radiative emission occurs. Regions with differing $n$ could be stacked either vertically or laterally next to each other. Figures 5b-d show time-resolved PL spectroscopy of an ensemble in solution, with Figure 5b showing that upon excitation, the $n=2$ fraction emits a very short burst of light that very quickly dies off, as its energy is rapidly transferred to higher $n$ fractions. This is reflected in the much longer $1/e$ decay time of the $n=4$ fraction, 88 ns, compared to the $n=2$ (150 ps) and $n=3$ fractions (140 ps), as shown in Figure 5c, d and inset. These observations agree with previous time-resolved spectroscopic studies for 2D perovskites.[28,29] The short DDA ligands used to prepare the MAPbBr sheets favored the energy funneling process. Synthesizing with longer TDA and HDA resulted in emission at higher energies (Figure S16), away from the bulk, confirming that the energy transfer process is much weaker. Longer octadecylamine (ODA) does not favor the formation of layered structures, and produces bulk nanosheets with a corresponding emission.

Energy funneling towards lower bandgap regions, where radiative recombination processes are more efficient, results in high PL quantum yields (PLQY) of up to 49% for the MAPbBr sample with $n=4$ main fraction. The high PLQY value is striking, considering that the volume of the bulk-like regions must be small, at most 10%, compared to that of the layered regions based on optical absorption and XRD data (minor bulk-like features at 15 and 30° can be deconvoluted from the XRD measurements for $n=4$, as shown in S17), and is evidence of the efficiency of the energy funneling process. Indeed, such nanograins of bulk MAPbBr have been engineered recently to enhance the efficiency of light emitting diodes by discouraging the dissociation of excitons into unbound charge carriers.[30] If energy funneling were not present, it would require that the bulk regions have PLQYs over 100%, considering their approximate volumes and that about 70% of the PL comes from these regions. The quantum yield is significantly lower for MAPbBr nanosheets with lower $n$ value, down to 11% for $n=2$ main fraction. Despite having an increased exciton binding energy and radiative recombination rate, charges in thinner layers are less effectively screened from surface defects, resulting in lower quantum yields.[9] The group of Tisdale *et al.* reported an increase of PLQY for colloidal 2D perovskites from 6% for MAPbBr to 22% for formamidinium-based analogue with n=2.[31] A PLQY reaching 70% was reported by the group of Urban *et al.* for mixed lead bromide-iodide perovskite nanoplatelets obtained by the exfoliation method.[14] Recently, quasi-2D perovskites in films on the basis of 5-aminovaleric acid cross-linked MAPbBr were presented with PLQY reaching 80%.[32] Thus our nanosheets present high quality and effective perovskites among solution-processable hybrid 2D MAPbBr nanostructures obtained by hot-injection colloidal synthesis offering additionally the potential of large lateral size.

An energy funneling process is observable in the MAPbI sample with $n=3$ as the main fraction, having a PLQY of 1%. The generally lower PLQY can be partly attributed to its lower exciton binding energy (about three times lower than MAPbBr).[33] In addition, energy funneling effects between regions of differing $n$ will be significantly hampered by the longer organic spacer separating them (2.2 nm for MAPbI compared to 1.7 nm for MAPbBr). Indeed, in order to achieve efficient energy funneling in MAPbI films, Yuan *et al.* used the relatively short phenylethylammonium (~0.8 nm) ligand. Similarly, the 2.5 nm organic spacer for the MAPbCl series is expected to hinder energy funneling to low bandgap regions, resulting in PL from all of the peaks in the nanosheets, as shown in Figure 5c. The organic spacer ligands, longer in this work compared to those employed in other studies, enhances the potential barrier around the perovskite layers, intensifying their quantum confinement. Whereas the $n$=1 MAPbI PL in this work is centered at 502 nm, it was at 512,[31] 528 and 539 nm for similar $n$=1 nanosheets with shorter spacer ligands.[6,14] These results show that the ligand choice for the formation of 2D perovskite NSs might have dramatic influence on their optical properties and should be taken into account during designing devices based on these materials. Additionally, investigations regarding a change in the PL with different lateral sizes showed no significant shift of the peak (Figure S18).

## CONCLUSION

In summary, highly tunable, colloidal route protocols are presented for synthesizing micron-sized nanosheets of MAPbX (X=I, Br, Cl) with 2D (layered, $n$=1 to 4) and 3D structure, as well as good control over both the lateral size and thickness. Excitonic features in the optical spectra confirmed the 2D nature of the nanosheets, and revealed an energy funneling process from low to high $n$ regions, mediated by the length of the organic ligands. This efficient process, which resulted in strong emission from the high $n$ regions (PLQY up to 49%) even if these were present only in very small quantities, could be used to engineer high performance optoelectronic devices.

## METHODS

**Chemicals and reagents.** All chemicals were used as received: Lead(II) acetate tri-hydrate (Aldrich, 99.999%), oleic acid (OA, Aldrich, 90%), nonanoic acid (Alfa Aesar, 97%), tri-octylphosphine (TOP; ABCR, 97%), 1,2-diiodoethane (DIE; Aldrich, 99%), methylammonium bromide (MAB; Aldrich, 98%), methylammonium chloride (MAC; Aldrich, 98%), methylammonium iodide (MAI; Aldrich, 98%), diphenyl ether (DPE; Aldrich, 99%), toluene (VWR, 99,5%), dimethylformamide (DMF; Aldrich, 99,8%), 1-bromotetradecane (BTD; Aldrich, 97%), 1-chlorotetradecane (CTD; Aldrich, 98%), octadecylamine (ODA; Aldrich, 97%), tetradecylamine (TDA; Aldrich, 95%), hexadecylamine (HDA; Aldrich, 90%), dodecylamine (DDA; Merck, 98%), oleylamine (ACROS, 80-90%).

**Synthesis of n>4 MAPbI$_3$ nanosheets.** *Standard synthesis:* A three neck 50 mL flask was used with a condenser, septum and thermocouple. 10.5 mL of diphenyl ether (66.6 mmol), 0.48 mL of a 400 mg hexadecylamine (1.66 mmol) in 4 mL diphenyl ether precursor and 0.1 mL (0.22 mmol) of TOP were heated to 80 °C in a nitrogen atmosphere. Then vacuum was applied to dry the solution. After 1 h the reaction apparatus was filled with nitrogen again and 2 mL of as prepared PbI$_2$ nanosheets in toluene were added. The reaction temperature was reduced to 35 °C after all of the PbI$_2$ dissolved. The synthesis was started with the injection of 0.06 mL of a 600 mg methylammonium iodide (3.77 mmol) in 6 mL dimethylformamide precursor. After the injection the temperature was slowly increased to 90 °C for a time period of 9 minutes. At 90 °C the heat source was removed and the solution was left to cool down below 60 °C. Afterwards, it was centrifuged at 4000 rpm for 3 minutes. The particles were washed two times in toluene before the product was finally suspended in toluene again and put into a freezer for storage.

**Synthesis of n>4 MAPbBr$_3$ nanosheets.** *Standard synthesis:* A three neck 50 mL flask was used with a condenser, septum and thermocouple. 10.5 mL of diphenyl ether (66.6 mmol), 0.06 mL of oleylamine (0.18 mmol) and 0.2 mL (0.44 mmol) of TOP were heated to 80 °C in a nitrogen atmosphere. Then vacuum was applied to dry the solution. After 1 h the reaction apparatus was filled with nitrogen again, the temperature was increased at 120 °C and 1.5 mL of as prepared PbBr$_2$ nanosheets in toluene were added. The reaction temperature was reduced to 35 °C after all of the PbBr$_2$ dissolved. The synthesis was started with the injection of 0.06 mL of a 300 mg methylammonium bromide (2.68 mmol) in 6 mL dimethylformamide precursor. After the injection the temperature was increased to 120 °C. After 10 minutes the heat source was removed and the solution was left to cool down below 60 °C. Afterwards, it was centrifuged at 4000 rpm for 3 minutes. The particles were washed two times in toluene before the product was finally suspended in toluene again and put into a freezer for storage.

**Synthesis of n>4 MAPbCl$_3$ nanosheets.** *Standard synthesis:* A three neck 50 mL flask was used with a condenser, septum and thermocouple. 10.5 mL of diphenyl ether (66.6 mmol), 0.04 mL of oleylamine (0.12 mmol) and 0.1 mL (0.22 mmol) of TOP were heated to 80 °C in a nitrogen atmosphere. Then vacuum was applied to dry the solution. After 1 h the reaction apparatus was filled with nitrogen again, the temperature was increased at 220 °C and 1 mL of as prepared PbCl$_2$ nanosheets in toluene was added. After all of the PbCl$_2$ was dissolved the temperature was reduced to 100 °C. The synthesis was started with the injection of 0.36 mL of a 50 mg methylammonium chloride (0.74 mmol) in 6 mL dimethylformamide precursor. After 10 minutes the heat source was removed and the solution was left to cool down below 60 °C. Afterwards, it was centrifuged at 4000 rpm for 3 minutes. The particles were washed two times in toluene before the product was finally suspended in toluene again and put into a freezer for storage.

**Synthesis of n<4 MAPbI$_3$ nanosheets.** *Standard synthesis:* A three neck 50 mL flask was used with a condenser, septum and thermocouple. 10.5 mL of diphenyl ether (66.6 mmol), 0.48 mL of a 400 mg hexadecylamine (1.66 mmol) in 4 mL diphenyl ether precursor and 0.1 mL (0.22 mmol) of TOP were heated to 80 °C in a nitrogen atmosphere. Then vacuum was applied to dry the solution. After 1 h the reaction apparatus was filled with nitrogen again and 2 mL of as prepared PbI$_2$ nanosheets in toluene were added. The reaction temperature was reduced to 35 °C after all of the PbI$_2$ dissolved. The synthesis was started with the injection of 0.06 mL of a 600 mg methylammonium iodide (3.77 mmol) in 6 mL dimethylformamide precursor. After the injection the temperature was slowly increased to 60 °C for a time period of 6 minutes. At 60 °C the heat source was removed and the solution was centrifuged at 4000 rpm for 3 minutes. The particles were washed two times in toluene before the product was finally suspended in toluene again and put into a freezer for storage.

**Synthesis of n<4 MAPbBr$_3$ nanosheets.** *Standard synthesis:* A three neck 50 mL flask was used with a condenser, septum and thermocouple. 10.5 mL of diphenyl ether (66.6 mmol), 0.2 mL of a 500 mg dodecylamine (2.70 mmol) in 4 mL diphenyl ether precursor and 0.2 mL (0.44 mmol) of TOP were heated to 80 °C in a nitrogen atmosphere. Then vacuum was applied to dry the solution. After 1 h the reaction apparatus was filled with nitrogen again, the temperature was increased to 160 °C and 2.5 mL of as prepared PbBr$_2$ nanosheets in toluene was added. The synthesis was started after all of the PbBr$_2$ dissolved with the injection of 0.03 mL of a 300 mg methylammonium bromide (2.68 mmol) in 6 mL dimethylformamide precursor. After 5 minutes the heat source was removed and the solution was left to cool down below 60 °C. Afterwards, it was centrifuged at 4000 rpm for 3 minutes. The particles were washed two times in toluene before the product was finally suspended in toluene again and put into a freezer for storage.

**Synthesis of n<4 MAPbCl$_3$ nanosheets.** *Standard synthesis:* A three neck 50 mL flask was used with a condenser, septum and thermocouple. 10.5 mL of diphenyl ether (66.6 mmol), 0.24 mL of a 400 mg hexadecylamine (1.66 mmol) in 4 mL diphenyl ether precursor and 0.1 mL (0.22 mmol) of TOP were heated to 80 °C in a nitrogen atmosphere. Then vacuum was applied to dry the solution. After 1 h the reaction apparatus was filled with nitrogen again, the temperature was increased to 220 °C and 1 mL of as prepared $PbCl_2$ nanosheets in toluene was added. The reaction temperature was reduced to 100 °C after all of the $PbI_2$ dissolved. The synthesis was started with the injection of 0.24 mL of a 50 mg methylammonium chloride (0.74 mmol) in 6 mL dimethylformamide precursor. After 10 minutes the heat source was removed and the solution was left to cool down below 60 °C. Afterwards, it was centrifuged at 4000 rpm for 3 minutes. The particles were washed two times in toluene before the product was finally suspended in toluene again and put into a freezer for storage.

Variation of the dimensions for all three materials for bulk and for the layered structures is described in detail in the Supporting Information.

**Characterization.** The TEM samples were prepared by diluting the nanosheet suspension with toluene followed by drop casting 10 µL of the suspension on a TEM copper grid coated with a carbon film. Standard images were done on a JEOL-1011 with a thermal emitter operated at an acceleration voltage of 100 kV. X-ray diffraction (XRD) measurements were performed on a Philips X'Pert System with a Bragg-Brentano geometry and a copper anode with a X-ray wavelength of 0.154 nm. The samples were measured by drop-casting the suspended nanosheets on a <911> or <711> cut silicon substrate. Atomic force microscopy (AFM) measurements were performed in tapping mode on a JPK Nano Wizard 3 AFM in contact mode. Images were taken of the as-prepared nanoring devices. UV/vis absorption spectra were obtained with a Cary 5000 spectrophotometer equipped with an integration-sphere. The PL spectra measurements were obtained by a fluorescence spectrometer (Fluoromax-4, Horiba Jobin Yvon). Simulations of XRD spectra were carried out in PowderCell 2.4 software using crystallographic data from the literature.[19-21] The structure was simplified excluding the organic part of each unit cell, thus simulating solely Pb-X (X=Cl, Br, I) networks. Absolute quantum yield measurements were performed in solution with K-Sphere 'Petite' Integrating Sphere (Horiba) and Fluorolog-3 with FluorEssence software. Liquid samples were prepared as low-concentrated nanosheets solutions in toluene (optical density in the range of 0.02–0.05) in quartz cuvettes (QG). Excitation and emission spectra of both QD solutions and pure toluene were recorded at the excitation wavelength of 420 nm and recalculated in photons absorbed and emitted by the nanosheets according to the absolute 4-step measurement method. Time-resolved PL measurements were performed with Picoquant FT300 fluorescence spectrometer (1200 lines/mm grating, 30 cm focal length, PMA Hybrid Detector). Excitation was carried out with SuperK FIANIUM FIA-15 white-light laser with LLTF contrast tunable single line filter (1.5 nm bandwidth). Pulse duration was set to 60 ps, excitation wavelength to - 410 nm.

## ASSOCIATED CONTENT

***Supporting Information**

The Supporting Information is available free of charge on the ACS Publications website at DOI:

Additional experimental details; optical absorbance spectra; emission spectra; characterization of XRDs performed in a capillary for bulk and layered perovskite NSs; SAED patterns; TEM images; UV/Vis and emission spectra of dissolved lead halide nanosheets; AFM images and measured height images of synthesized perovskite nanosheets; TEM analysis and XRD patterns; difference in lateral dimensions and thickness; TEM images of pyramidal nanosheets; TEM images and SAED patterns of aliquots taken during a perovskite synthesis.

## ACKNOWLEDGMENTS

The authors thank the Alf Mews group for providing the Confocal Microscopy setup. Further, the German Research Foundation DFG is acknowledged for financial support in the frame of the Cluster of Excellence "Center of ultrafast imaging CUI" and for granting the project KL 1453/9-2. The European Research Council is acknowledged for funding an ERC Starting Grant (Project: 2D-SYNETRA (304980), Seventh Framework Program FP7). We further acknowledge MINECO (Spain) for the project MAT2016-81118-P.

**TABLE OF CONTENTS**

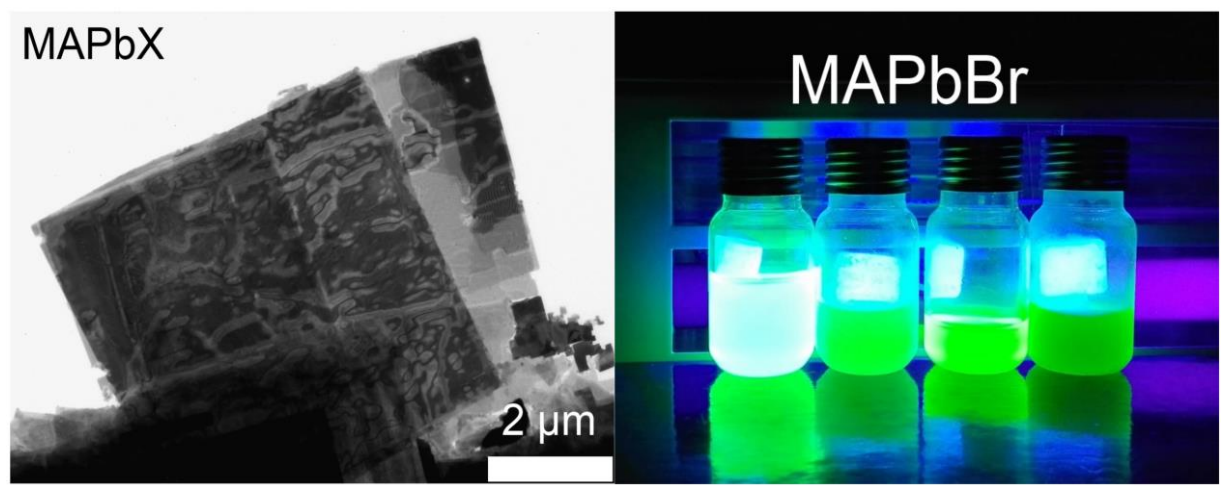